\documentclass[12pt,aps,pra,reprint,superscriptaddress,amsmath,amsfonts,amssymb,longbibliography,onecolumn]{revtex4-2}
\usepackage[T1]{fontenc}

\usepackage[most]{tcolorbox}
\newtcolorbox{myframe}[1][]{
  colframe=white,
  arc=0pt,
  outer arc=0pt,
  colback=white,
  boxrule=0.pt,
  #1
}

\usepackage{graphicx}% Include figure files
\usepackage{multirow,dcolumn}% Align table columns on decimal point
\usepackage{bm}% bold math
\usepackage[colorlinks=true,citecolor=blue,linkcolor=blue, urlcolor=blue]{hyperref}
\usepackage[acronym,shortcuts]{glossaries}

\usepackage{lipsum}
\usepackage{braket}

\usepackage{todonotes}

\begin{document}
\title{Open Hardware Solutions in Quantum Technology
%Open Quantum Hardware %for Collaborative Innovation
}
% Open Hardware in Quantum Technology for Collaborative Innovation

% Open Hardware in Quantum Technology: From scientific discovery to Collaborative Innovation


\author{Nathan Shammah}
\email{Email: nathan@unitary.fund}
\affiliation{Unitary Fund, San Francisco,
CA 94104, USA}

\author{Anurag Saha Roy}
%\email{Email: anurag@qruise.com}
\affiliation{Qruise GmbH, 66113, Saarbruecken, Germany}
% \affiliation{Forschungszentrum Jülich}
% \affiliation{Saarland University}

\author{Carmen G. Almudever}
%\email{Email: cargara2@disca.upv.es}
\affiliation{Technical University of Valencia, Spain}

\author{Sébastien Bourdeauducq}
%\email{Email: sb@m-labs.hk}
\affiliation{M-Labs Limited, North Point, Hong Kong}

\author{Anastasiia Butko}
%\email{Email: }
\affiliation{Applied Mathematics and Computational Research Division, Lawrence Berkeley National Laboratory, Berkeley, CA 94720, USA}

\author{Gustavo Cancelo}
\affiliation{Fermi National Accelerator Laboratory, Batavia, IL 60510, USA}

\author{Susan M. Clark}
%\email{Email: }
\affiliation{Sandia National Laboratories, Albuquerque, New Mexico 87123, USA}

\author{Johannes Heinsoo}
%\email{Email: johannes@meetiqm.com}
\affiliation{IQM Quantum Computers, Espoo 02150, Finland}

\author{Loïc Henriet}
%\email{Email: loic@pasqal.com}
\affiliation{PASQAL, 7 rue Leonard de Vinci, 91300 Massy}

\author{Gang Huang}
%\email{Email: }
\affiliation{Accelerator Technology and Applied Physics Division, Lawrence Berkeley National Laboratory, Berkeley, CA 94720, USA}

\author{Christophe Jurczak}
\affiliation{Quantonation, 75010 Paris, France}

\author{Janne Kotilahti}
%\email{Email: janne@meetiqm.com}
\affiliation{IQM Quantum Computers, Espoo 02150, Finland}

\author{Alessandro Landra}
%\email{Email: alessnadro@meetiqm.com}
\affiliation{IQM Quantum Computers, Espoo 02150, Finland}

\author{Ryan LaRose}
%\email{Email: rlarose@umich.edu}
%\affiliation{Unitary Fund}
\affiliation{Center for Quantum Computing, Science, and Engineering, Michigan State University [East Lansing, MI 48824, USA}

\author{Andrea Mari}
%\email{Email: andrea@unitary.fund }
\affiliation{Unitary Fund, San Francisco,
CA 94104, USA}
\affiliation{School of Science and Technology, Universit\`a di Camerino, 62032 Camerino, Italy}

\author{Kasra Nowrouzi}
%\email{Email: }
\affiliation{Applied Mathematics and Computational Research Division, Lawrence Berkeley National Laboratory, Berkeley, CA 94720, USA}

\author{Caspar Ockeloen-Korppi}
%\email{Email: caspar@meetiqm.com}
\affiliation{IQM Quantum Computers, Espoo 02150, Finland}

\author{Guen Prawiroatmodjo}
%\email{Email: }
\affiliation{Microsoft Quantum, Redmond, Washington 98052, USA}

\author{Irfan Siddiqi}
%\email{Email: }
\affiliation{Quantum Nanoelectronics Laboratory, Department of Physics, University of California at Berkeley, Berkeley, CA 94720, USA}
\affiliation{Applied Mathematics and Computational Research Division, Lawrence Berkeley National Laboratory, Berkeley, CA 94720, USA}

%\author{Gary A. Steele}
%\email{Email: g.a.steele@tudelft.nl}
%\affiliation{Kavli Institute of Nanoscience, Delft University of Technology, 2628 CJ Delft, The Netherlands}

\author{William J. Zeng}
%\email{Email: will@unitary.fund}
\affiliation{Unitary Fund, San Francisco,
CA 94104, USA}
\affiliation{Quantonation, 75010 Paris, France}

%\author{Authors}
%%\email{Email: }
%\affiliation{Institution}

\begin{abstract}
Quantum technologies such as communications, computing, and sensing offer vast opportunities for advanced research and development. While an open-source ethos currently exists within some quantum technologies, especially in quantum computer programming, we argue that there are additional advantages in developing open quantum hardware (OQH). Open quantum hardware encompasses open-source software for the control of quantum devices in labs, blueprints and open-source toolkits for chip design and other hardware components, as well as openly-accessible testbeds and facilities that allow cloud-access to a wider scientific community. We provide an overview of current projects in the OQH ecosystem, identify gaps, and make recommendations on how to close them today. More open quantum hardware would accelerate technology transfer to and growth of the quantum industry and increase accessibility in science.
\end{abstract}

\maketitle

%\tableofcontents
\section{Introduction}
 The free and open exchange of scientific tools has become more and more important as automation and devices increase their role into scientific fields.
The past five years have witnessed an explosion of open-source tools for programming quantum computers. The open nature of these software tools has substantially increased the number of users of quantum computers, and created a new genre of programmer: the ``quantum software engineer''~\cite{Zeng_2017,Fingerhuth_2019,Monroe_2019,Fox21,Asfaw22,Dzurak22}. This has shaped the development of quantum computing as a whole, leading to the creation of new organizations, job categories, and career paths.

\begin{figure*}
    \centering
    \includegraphics[width=0.5\textwidth]{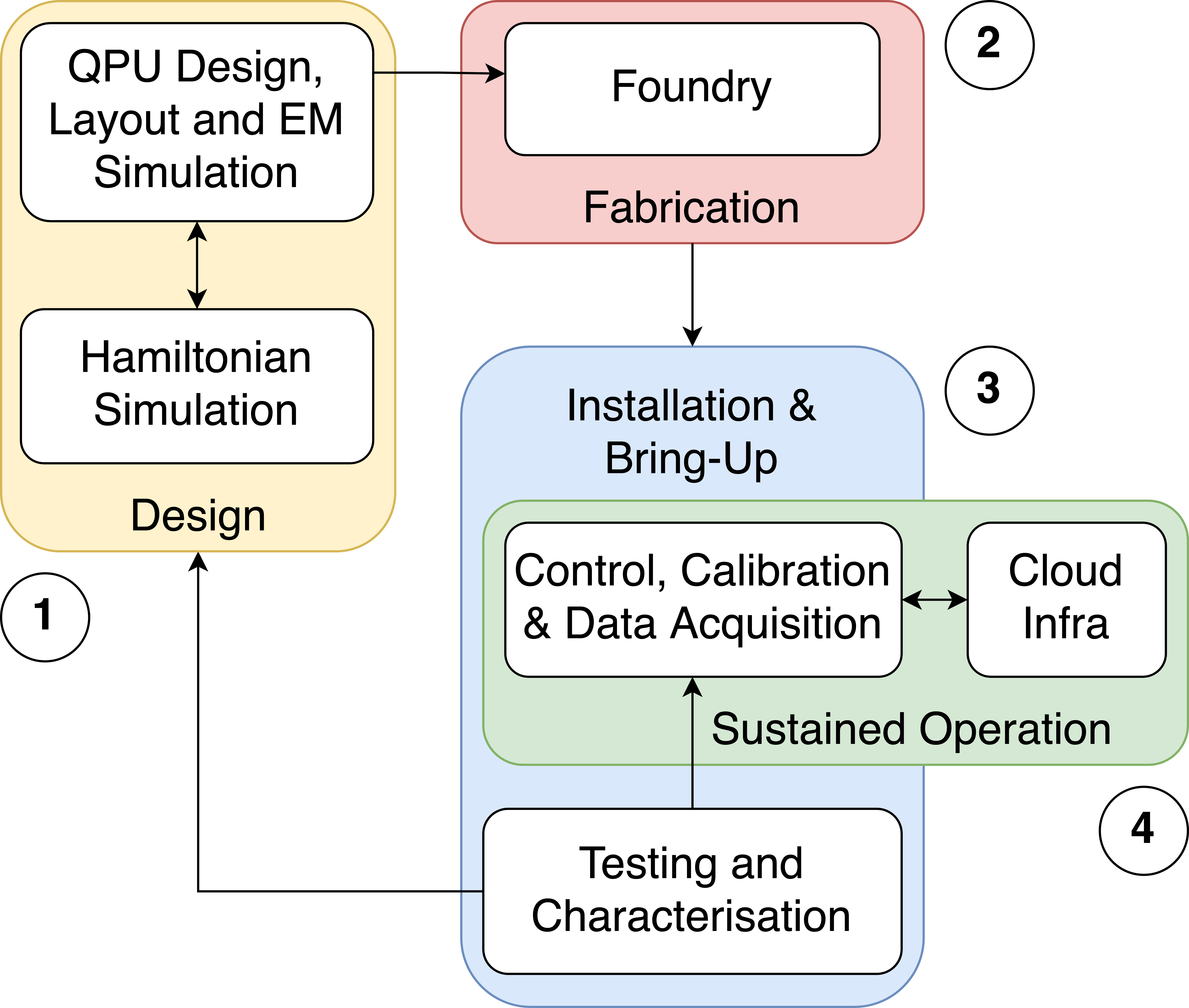}
    \caption{{\bf{Overarching diagram of the Open Quantum Hardware steps and their interconnection}}. 1: Design phase, which can involve a loop between simulation of the Hamiltonian and electromagnetic (EM) simulations used to define the QPU design and layout. 2: Fabrication step, e.g., through a foundry. 3: Installation and bring-up, which includes testing and characterization; information collected at this step can be fed back to step 1 to modify designs. 4: Sustained operation, with overlaps (with step 3), and is composed of data acquisition (with control and calibration) and can involve interfacing with infrastructure to provide cloud access to the device or experiment. While these steps appear also in more general quantum hardware projects that are not open, their separation and standardization is a required feature to interconnect open frameworks.}
    \label{fig:overarching-diagram}
\end{figure*}

Less attention has been paid to the tools and components actually used to build and control quantum computers -- and quantum technologies such as communications and sensing more broadly -- as well as efforts making quantum computers more accessible in non-commercial ways. We use the term ``open quantum hardware'' (OQH) to cover them, and intend for it to explicitly encompass several steps related to the openness associated to hardware in quantum technology, including: (1) open-source software (OSS) for designing quantum processors and other hardware components (used for computation, but also for simulation, sensing, and communication), (2) foundries and facilities for fabricating quantum processors, (3) software for controlling, analyzing and characterizing quantum devices, and (4) software and hardware for making the infrastructure that enables various levels of open-access -- from cloud-accessible quantum processors to collaborative testbeds providing non-commercial access and testing, from remotely accessible research labs to the related cloud infrastructure. These steps covering the life cycle of OQH and their overlapping relations are sketched in Fig.~\ref{fig:overarching-diagram}. 

In this article, we provide an overview of efforts in each of these categories of open quantum hardware (OQH), highlight notable projects in each category, identify gaps, and provide recommendations for closing them. A major theme in this review is that there is much opportunity to promote interoperability, reduce cost, and increase the number of users of quantum technologies.

Enabling the open quantum hardware ecosystem is both the natural complement to the open quantum software ecosystem (which is itself quite robust and sophisticated~\cite{Fingerhuth_2019}), as well as the natural extension of efforts to open up classical hardware. This ecosystem also represents a prime opportunity for entities building hardware (of all kinds) to engage in the kind of beneficial, pre-competitive activity which boosts the ecosystem as a whole (and consequently, those builders themselves). 

A first generation of projects in open quantum hardware, such as ARTIQ and pyEPR~\cite{Bourdeauducq_2016,Minev21}, has pioneered a free and open dissemination of tools. We are now witnessing the beginning of a more robust and sophisticated OQH ecosystem~\cite{Scholten20,Shammah21ws} which can enable and accelerate scientific progress and discovery on a wider scale. On the one hand, this ecosystem can further extend beyond quantum computing, on which it is mostly focused, to encompass quantum technologies such as quantum sensing and communications. On the other hand, it can further integrate and benefit from the adoption and integration of existing non-quantum open hardware projects and frameworks \cite{klayout}.
What's more, given the substantial investments made around the world~\cite{thew2019focus} -- especially in Australia~\cite{roberson2019charting}, Eurasia~\cite{zhang2019quantum,riedel2019europe,gibney2017billion,knight2019uk,yamamoto2019quantum,fedorov2019quantum}, and North America~\cite{Merzbacher_2020,raymer2019us, monroe2019us,sussman2019quantum} -- to support the development of a quantum technologies industry, ensuring those investments yield the best possible fruit is of great importance. One way to do so is through encouraging the development of open communities and ecosystems~\cite{Aiello21,Mirowski18,Powell12,Viseur12,Morreale16,Morreale17,Blind21,Soriano14,Virtanen20,Harris20} around quantum technology.

%These tools and components include those for electronic design and analysis, control systems, and the “bare metal” of qubit design itself. Further, efforts are underway to increase the accessibility of quantum computers through testbed initiatives in the United States at the federal level through the involvement of national labs. In the European Union, this is happening through the HPCQS program (https://www.hpcqs.eu/) as well as in some national initiatives, such as in Sweden with the work ongoing at Chalmers and in the WAQCT [Citation needed]. These schemes provide end-users of quantum computers access in a non-commercial way that can be beneficial to foster progress in the scientific space.

To describe the potential of OQH, we can look at the open quantum software ecosystem, which has flourished within the past five years, thanks to seeds planted over ten years ago. As a result, researchers (theorists and experimentalists alike) as well as quantum software engineers worldwide can use open-source software to advance quantum technology and quantum science in many directions, generally building upon an existing stack. An illustrative case is that of the Quantum Toolbox in Python (QuTiP)~\cite{Johansson_2012,qutip2}, first released in 2012, which enables the exploration of the effects of noise on a variety of quantum systems interacting with the environment. Building on top of QuTiP, several other tools have emerged, focusing on specific niches such as nontrivial system-environment dynamics~\cite{Shammah18,Lambert_2019}, notably getting closer and closer to the simulation of QPUs and their diagnostics~\cite{Groszkowski21,Li2022}. 

When looking at similar efforts in the classical open hardware space, the astounding success of the Arduino project~\cite{Banzi} (microcontrollers) as well as the Raspberry Pi ~\cite{richardson2012getting} (single-board computers) speaks to the power of putting open hardware in the hands of end-users. We can reasonably expect similar benefits within the quantum hardware ecosystem. Compared to only open-source software, an open hardware ecosystem also provides legibility, transparency, and reproducibility of hardware devices, a crucial consideration for supply chains and their management.

%When looking at similar efforts in the classical hardware space, the astounding success of the Arduino project (microcontrollers) as well as the Raspberry Pi (single-board computers) speaks to the power of putting open hardware in the hands of end-users.

Finally, the pre-competitive activity which can take place by opening up quantum hardware extends these benefits by mobilizing institutional momentum from key players who can help the ecosystem adopt a broad ethos of openness, interchangeability, interoperability, and benchmarking, as fostered by bodies like the Quantum Economic Development Consortium (QED-C) in the USA and similar entities in other geographies~\cite{Lubinski2023,Lubinski2023optimization,Amico2023}. In doing so, this allows for additional adoption and use of quantum hardware in academic labs and testbeds (a point returned to in Sec.~\ref{sec:testb}).

This article is organized as follows: Sec.~\ref{sec:open-quantum-hardware-today} gives a high-level overview of the state of the art in OQH, divided among blueprints and software for hardware design (Sec.~\ref{sec:HWdesign}), and software for control and data acquisition (Sec.~\ref{sec:control-and-data-acquisition}), itself divided into data acquisition (\S~\ref{sec:data}), pulse-level control on hardware (\S~\ref{sec:hardware}), pulse-level simulation (\S~\ref{sec:pulse}), and optimal control, calibration and characterization (\S~\ref{sec:control}). 
In Sec.~\ref{sec:qec} we provide an outlook on the current status and specific need for OQH for quantum error correction. 
In Sec.~\ref{sec:facilities} we review facilities for OQH, such as remotely accessible labs (Sec.~\ref{sec:remotelabs}), Testbeds (\S~\ref{sec:testb}) and Foundries (\S~\ref{sec:foundr}). Throughout Sec.~\ref{sec:open-quantum-hardware-today}'s subsections, we focus on some example projects to bring definiteness to the discussion. 
In Section \ref{sec:gaps} we discuss gaps in OQH and make recommendations to close them. Finally, in Section \ref{sec:concl}, we give our conclusions.

\section{Open quantum hardware today} \label{sec:open-quantum-hardware-today}
Open quantum hardware encompasses open-source software that is used for designing, analyzing, building, controlling, and programming quantum chips, foundries which build quantum chips, and cloud-accessible labs and testbeds that provide alternative, non-commercially-driven access to them. In this section, we review the state of open quantum hardware along each of these categories. 
It should be noted that within each of these categories, there can exist multiple stacks -- configurations of components which are all inter-operable with one another, but not necessarily with components used in a different stack. Ensuring inter-operability of components across stacks would help ensure a more robust and sophisticated ecosystem. 
In Table~\ref{table:1} we summarize the different categories, providing examples of existing open hardware projects and facilities. We group them by functionalities: different open-source and open-hardware projects are present in processor design, simulation and diagnostics, control and data acquisition. We also group facilities and organizations: remotely accesible labs, testbeds, and foundries.

In Table~\ref{table:1b}, we highlight how different projects and facilities (emphasized in italics in the table) are distributed on the main quantum technology architectures: atoms, ions, photonics, spins and SC qubits. It is visible that some architectures are populated with more projects, such as SC qubits and atoms. On the contrary, just a few tools exist for ions, with notable exceptions provided by ARTIQ and the Open Quantum Design foundation. Tools that have a strong cross-platform adoption are shown in the last row of Table~\ref{table:1b}. In the following sections we explore in greater depth how different tools have been developed for different functionalities and are used in various architectures.

\begin{table}[t!]
\centering
\def\arraystretch{1.8}%  1 is the default, change whatever you need
\setlength\tabcolsep{.2cm}

\begin{tabular}{p{0.03\textwidth}p{0.25\textwidth}p{0.5\textwidth}}
%\hline & & \\
& {\bf Functionality}& {\bf Examples }  \\
%& \\
\hline
\hline
%&  \\
\parbox[t]{2mm}{\multirow{6}{*}{\rotatebox[origin=c]{90}{{\bf Projects}}}}
&Processor Design & DASQA~\cite{dasqa, kunasaikaran2023framework}, KQCircuits~\cite{kqcircuits},  PainterQubits/Devices.jl, pyEPR~\cite{Minev21}, Qiskit Metal~\cite{Qiskit_Metal}, QuCAT~\cite{Gely_2020}    \\
%&   \\
%&   \\
&Simulation and diagnostics &
KQCircuits~\cite{kqcircuits}, Pulser~\cite{Silverio_2022}, Qiskit Metal~\cite{Qiskit_Metal},  QuTiP~\cite{Johansson_2012,qutip2}, QuTiP-QIP~\cite{Li2022}, sc-qubits~\cite{Groszkowski21}, Strawberry Fields~\cite{Killoran_2019} %&Early on, finite element software was closed and heavily license restricted, rencently more open-source EM solvers are open. %all finite element software currently used fully closed and heavily license restricted 
\\    
%& \\
&Control and data acquisition &
ARTIQ~\cite{Bourdeauducq_2016}, Duke-ARTIQ~\cite{Duke-artiq}, Qua\footnote{partially open-source}~\cite{ella2023quantumclassical}, QCoDeS~\cite{jens_hedegaard_nielsen_2023_8344579}, QICK~\cite{Stefanazzi21}, Quantify~\cite{mesman2023qprofile}, QubiC~\cite{Xu2022}, Qudi~\cite{Qudi_Softwarex_2017}
qupulse~\cite{qupulse}, Sinara Open Hardware~\cite{Sinara,Kulik_2022}%&LabView, Labber (Keysight), Zurich Instruments LabOne, custom-made &Mixed, 
%& Field is generally closed but with a trend towards more openness
%\\
%&   \\
%Cloud infrastructure &
%IBM Quantum, Azure Quantum, AWS Braket, Rigetti Cloud, Quantum Inspire  
%&Some open
\\
\hline
\parbox[t]{2mm}{\multirow{5}{*}{\rotatebox[origin=c]{90}{{\bf Facilities}}}}
&Remotely Accessible Labs\footnote{excluding commercial providers} &
Forschungszentrum J\"ulich through OpenSuperQ \cite{opensuperq}, Quantum Inspire \cite{quantum-inspire}
\\

%&   \\

&Testing (Testbeds) &
Lawrence Berkeley National Lab's AQT, Open Quantum Design, Sandia National Labs' QSCOUT~\cite{Clark21}, Sherbrooke's Distriq DevTeQ, NQCC \cite{nqcc} 
%& Mostly closed, currently in academic clean rooms or industrial complexes\\
%&   
\\
&Fabrication (Foundries) &
LPS Qubit Collaboratory, UCSB quantum foundry, QuantWare\footnote{private company with support for Qiskit Metal} %&Typical foundries are mostly closed
\\

\hline
\hline

\end{tabular}
\caption{{\bf OQH examples grouped by functionality}. The different functionalities in open quantum hardware and examples of representative projects: Processor design, \S\ref{sec:HWdesign}; simulation and diagnostics, \S\ref{sec:control-and-data-acquisition}; and control and data acquisition, \S\ref{sec:control}. We also list programs related to facilites in OQH: Remotely accessible labs (where here we exclude here the many commercial providers and cloud services, that are discussed in \S\ref{sec:remotelabs}); testing (testbeds), \S\ref{sec:testb}; and fabrication (foundries), \S\ref{sec:foundr}.}
\label{table:1}
\end{table}

\begin{table}[ht!]
\centering
\def\arraystretch{2.5}%  1 is the default, change whatever you need
\setlength\tabcolsep{.5cm}

\begin{tabular}{p{0.09\textwidth}p{0.1\textwidth}p{0.1\textwidth}p{0.08\textwidth}p{0.1\textwidth}p{0.2\textwidth}}
%\hline & & \\
& {\bf Atoms
 }& {\bf Ions
 }& {\bf Photonics
 }& {\bf Spins
 }&{\bf SC qubits
}  \\
%& \\
\hline
\hline
%&  \\
{\bf Projects} &Bloquade.jl, Labscript-qc, Pulser, QICK & ARTIQ & Perceval,
Strawberry Fields
& QICK, Qudi, Qupulse
 &
ARTIQ, DASQA, KQCircuits, PyEPR, QICK, Qiskit Metal, QubiC, Sc-qubits
    \\
{\bf Facilities}&\emph{AQT}, \emph{OpenSuperQ}& \emph{Open Quantum Design}, \emph{QSCOUT}& & \emph{LPS Collaboratory}& 
\emph{LPS Collaboratory, Quantum Inspire}

    \\
\hline
\hline
{\bf Architecture-agnostic}&\multicolumn{5}{c}{QCoDeS, Qibolab, Qua, Quantify, QuOCS, QuTiP-QIP, \emph{QCUP}}\\     
\hline
\hline

\end{tabular}
\caption{{\bf OQH examples grouped by architecture}. The different open hardware projects, grouped by architectures (atoms, ions, photonics, spins, and SC qubits) The bottom row includes projects that are architecture-agnostics. Italics is used to highlight programs at facilities (e.g., instead of software projects).}
\label{table:1b}
\end{table}

\subsection{Blueprints and software for hardware design}
\label{sec:HWdesign}

Blueprints for hardware design are classic examples of ``open hardware''~\cite{Gao21,Tahan21,Fu17,Feng21,Castellanos_Gomez_2014,Ziegler_2022,Pollice21,Ma20,Molesky18,kunasaikaran2023framework}. In modern device design, software tools have been developed to tackle various aspects of this task, known as computer-aided design (CAD). In quantum computing research, while quantum device design is often published in peer-reviewed papers and micrographs are included in the supporting materials and figures, generally, CAD drawings are not shared in the open. Notable exceptions include pyEPR~\cite{Minev21}, Qiskit Metal~\cite{Qiskit_Metal}, and KQCircuits~\cite{kqcircuits}, three projects enabling various capabilities for chip design of superconducting-circuit (SC) qubits. 

Superconducting circuits incorporating non-linear devices, such as Josephson junctions and nanowires, are among the leading platforms for emerging quantum technologies. Using pyEPR, one can design and optimize SC circuits and control dissipative Hamiltonain parameters in a systematic, simple and robust approach. This reduces the number of required ab-initio simulations. pyEPR has been used on a variety of circuit quantum electrodynamics (cQED) devices and architectures, from 2D to 3D, including ``2.5D" (flip-chip), demonstrating 1\% to 10\% agreement for non-linear coupling and modal Hamiltonian parameters over five-orders of magnitude and across a dozen samples. 
 
\begin{figure}
    \centering
    \includegraphics[width=0.4\textwidth]{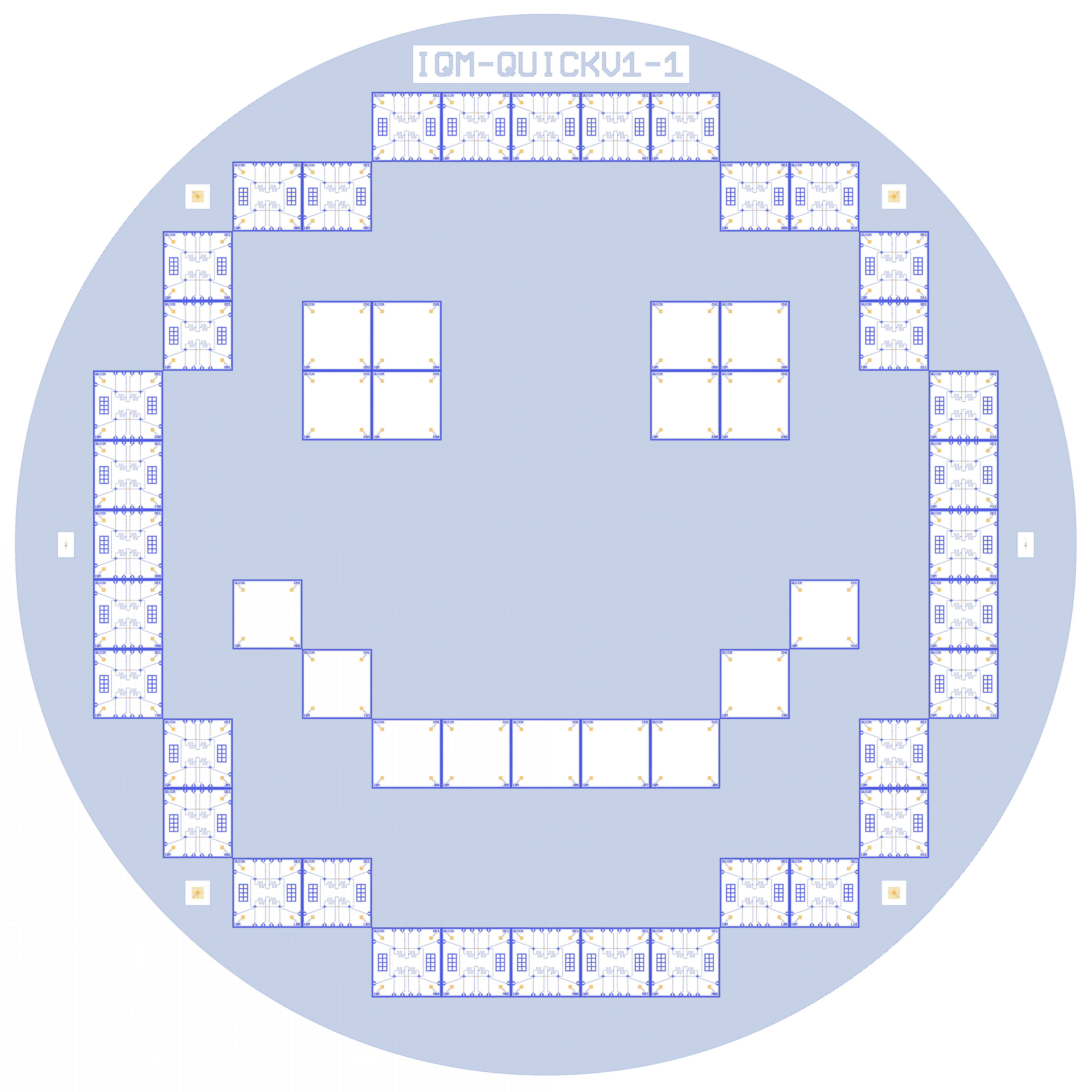}
    \includegraphics[width=0.4\textwidth]{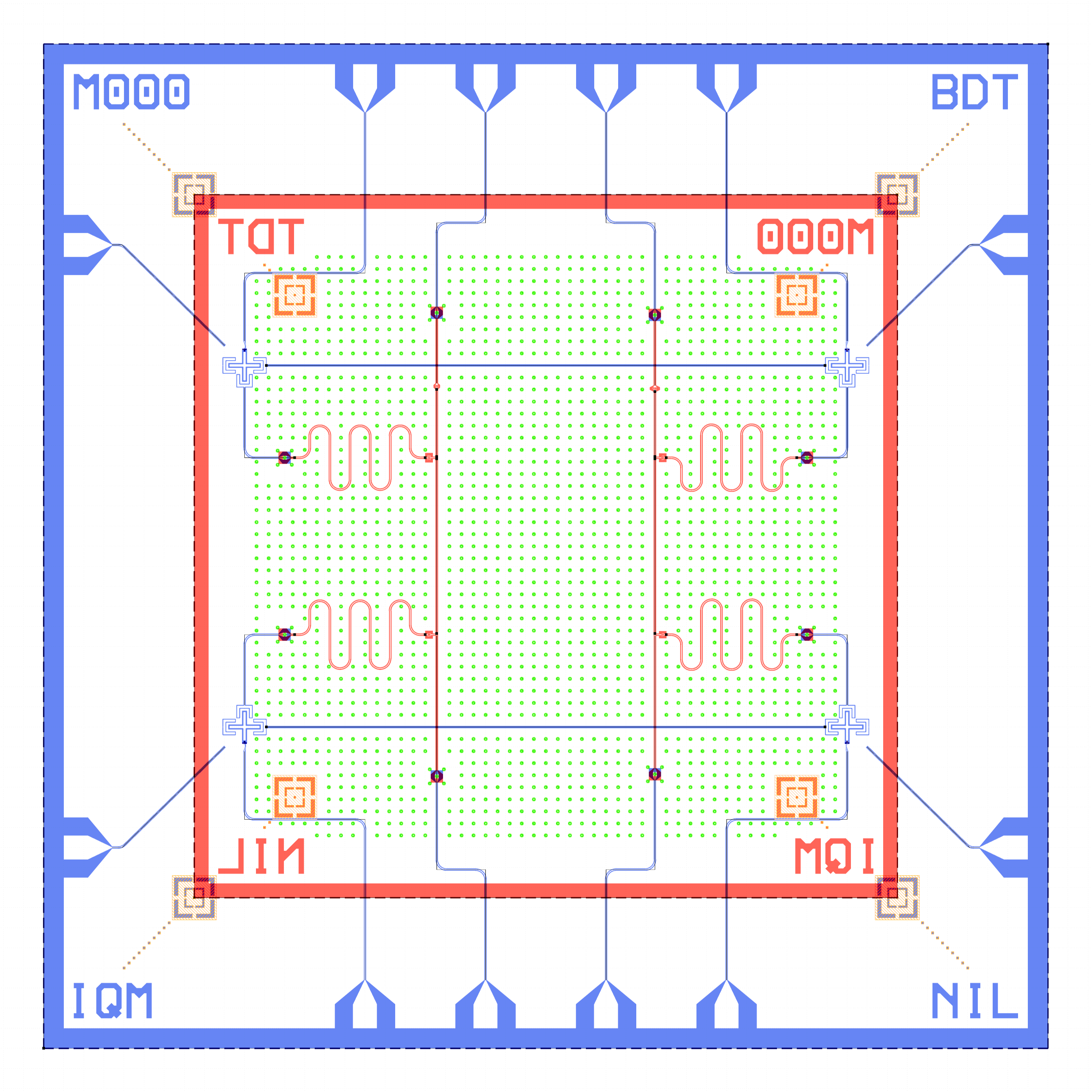}
    \caption{\textbf{Photomask layout and chip design.} An open-source photomask (left) provided in KQCircuits as a demo example. An individual SC quantum processor layout (right) based on flip chip technology, where blue is one substrate, red is the other substrate and in green denotes the bump bonds. The chip contains four transmons capacitively coupled and two buses for multiplexed readout. The code for the mask can be accessed here \href{https://github.com/iqm-finland/KQCircuits/blob/main/klayout_package/python/scripts/masks/quick_demo.py}{here} and for the chip \href{https://github.com/iqm-finland/KQCircuits/blob/main/klayout_package/python/kqcircuits/chips/demo_twoface.py}{here}.}
    \label{fig:demo_mask}
\end{figure}

Finite Element Methods (FEM) simulations, 
%as those shown in Figure~\ref{fig:pyepr-visualisation}, 
can be obtained with PyEPR using the  energy participation ratio approach (EPR). EPR unifies the design of dissipation and Hamiltonians around a single concept — the energy participation, a number between zero and one — in a single-step electromagnetic simulation. After the FEM simulations have validated the qubit properties (and possibly its coupling to a cavity), one is ready to fabricate the SC QPU. To this end, a photomask is generated to place the SC qubits on the substrate of a processor. 
You can find a typical CAD drawing of a superconducting flip chip QPU and an example mask generated by KQCircuits~\cite{kqcircuits} in Figure ~\ref{fig:demo_mask} and more information about KQCircuits in the frame below (\emph{Example 1}).
%\begin{figure}[ht]
%    \centering
%    \includegraphics[width=0.8\textwidth]{pyepr-vis.png}
%    \caption{{\bf Finite Element Methods (FEM) chip simulations} Design and FEM simulations of a cavity (left) and qubit (right) modes, using PyEPR~\cite{Minev21}. The contour plot of cavity modes displays the intensity of the electric field at 9.0 GHz. The current-density magnitude of a transmon qubit, showing transmon pads connected by a  Josephson junction [adapted from Ref.~\cite{Minev21}].}
%    \label{fig:pyepr-visualisation}
%\end{figure}
\begin{myframe}[width=\columnwidth]
\emph{Example 1, Hardware Design: }%\newline
{\bf ``KQCircuits: KLayout Python library for integrated quantum circuit design.'' }%\newline
KQCircuits~\cite{kqcircuits} is an open-source Python library created by IQM, a full-stack quantum computing startup, for designing superconducting quantum circuits. It automates the layout and simulation part of the design process, outputting chip layouts and photomask files in OASIS or GDSII formats, standard formats for the specification of data structures of photomasks, and integrated circuit layout. The layout files are sent to a mask manufacturer to produce a physical mask which is then used for the fabrication process. As a part of the mask export process, KQCircuits also produces other files such as netlists, to help with design verification. While KQCircuits itself cannot be used to perform simulations, it can be used to export automated simulation scripts and files for popular electromagnetic field simulation toolkits, such as Ansys HFSS/Q3D~\cite{ansys}, pyEPR~\cite{Minev21} (Energy Participation Ratio simulation), Sonnet~\cite{sonnet}, and Elmer~\cite{elmer}. These simulations use the parameterized geometry as the physical mask layouts produced by KQCircuits. A comparison between KQCircuits and Qiskit Metal is provided in Table \ref{table:2}.

KQCircuits generates multi-layer 2-dimensional-geometry representing common structures in quantum processing units (QPU). An important part of KQCircuits are the definitions of parameterized geometrical objects, or ``elements'', and a framework for easily defining new ones. Generating the designs using code makes it easy to quickly create different variations of designs and helps to avoid costly human errors. The combination of many elements into a full QPU design is made easier by features such as named reference points used for automatic positioning of elements relative to each other and Graphical User Interface (GUI) editing. Furthermore, KQCircuits includes a library of premade chips, many of which have been manufactured and used for testing at IQM.

KQCircuits works on top of KLayout~\cite{klayout}, an open-source layout design software that is mainly used for classical electronics. Advantageously, existing KLayout functionalities can be used for quantum hardware other than classical, while KQCircuits only needs to add features specific to QPU design. This connection with KLayout can also help to bring together the wider electronics open-source hardware community and open quantum hardware community. 
\end{myframe}

%\lstinputlisting[language=Python, firstline=25]{quick_demo.py}

In terms of remaining challenges for chip design tools, we note that computational electromagnetic simulation of quantum circuits is still heavily reliant on the use of proprietary and expensive tools such as Ansys HFSS or Sonnet or CST Studio (full-wave and capacitance simulations in particular). Open-source tools such openEMS or MEEP or Scuff-EM which are used elsewhere in the field of microwave simulations find very little adoption and application in quantum device design. Notable open-source exceptions are Elmer~\cite{elmer} and Palace~\cite{palace}. Elmer is an open-source parallel multi-physics Finite Element Methods (FEM) software used for quantum device simulation on desktop and High Performance Computing (HPC), e.g., 3D layouts capacitance matrices, cross section of layouts, London equations, which in the framework of OpenSuperQPlus (European Open-Access Quantum Computer Project \cite{opensuperq}) is developed with the partnership between CSC and IQM. Elmer along with Gmsh~\cite{gmsh} has also been integrated as a backend for mesh generation and FEM simulation in Qiskit Metal. Palace (Parallel, Large-scale Computational Electromagnetics)~\cite{palace}, is an open-source parallel FEM software capable of full-wave electromagnetics simulations developed at AWS, with out-of-the-box support for use of large scale cloud HPC resources.

\begin{table}[t]
\centering
\def\arraystretch{2}% 
\setlength\tabcolsep{.2cm}
\begin{tabular}{p{0.1\columnwidth}p{0.2\columnwidth}p{0.2\columnwidth}p{0.2\columnwidth}p{0.2\columnwidth}}
%\begin{tabular}{p{0.3\columnwidth}p{0.4\columnwidth}p{0.4\columnwidth}p{0.3\columnwidth}p{0.4\columnwidth}}
\hline%\\
{\bf Software}& {\bf Purpose}  & {\bf  Simulations}& {\bf Circuit analysis}& {\bf Input/output (I/O)} \\
\hline
%%& & & \\
Qiskit Metal & 
Full-stack quantum processor design.

& Ansys HFSS/Q3D, Ansys, GDS, etc. by plugins. Work in progress on open-source EM renders.
& EPR, impedance, quasi-lumped LOM, lumped.
& Python based.
Connects to Ansys, GDS, etc. by plugins. Work in progress on open-source EM renders.
\\
KQCircuits
 & 
Design parametrization and automation for quantum processors. Extensive components, chips and masks library.

& 
Ansys HFSS/Q3D, Sonnet, Elmer (open-source) supported.
& 
EPR, cross sectional EM properties, London equations and netlist export.
& 
Import/export GDSII and OASIS format. Netlist export. Automated simulation scripts and files export.

\\
%& & & \\
\hline

\end{tabular}
\caption{{\bf Open hardware design of SC qubit processors}. Qiskit Metal, supported by IBM, and KQCircuits, developed at IQM, are two software packages for open hardware design for SC qubit processors. We highlight their purpose, integration with FEM simulation software, circuit analysis characteristics and input and output file formats. EPR: Energy Participation Ratio, GDS: Graphic Design System, HFSS: High-Frequency Structure Simulator, LOM: Lumped Oscillator Model.}
\label{table:2}
\end{table}

\subsection{Control and data acquisition} \label{sec:control-and-data-acquisition}

In this section, we discuss the features and components involved in the execution and readout of quantum experiments. This field is not easily defined, as the operation of a QPU or quantum technology experiment is informed already at the highest level with the definition of abstract instructions, e.g., for quantum programming, gates in the form of a quantum circuit in a high-level software development kit (SDK). To narrow the field, we focus here on control that gets ``closer to the metal", e.g., lower than gate-level quantum circuit compilation in quantum computing. This field is already populated by a number of projects, spanning from pulse-level simulation of devices for diagnostics, characterization and calibration, to pulse-level control and data acquisition during experiments runs ~\cite{Alexander_2020,Ball20,Silverio2022,Li2022,Johansson_2012,Wittler_2021,roy2022software,jax2018github,Bourdeauducq_2016,Stefanazzi21,Groszkowski21,Xu21,jens_hedegaard_nielsen_2023_8344579,Li21,Roch20,Krenn21,Melnikov18,Smith22,Perkel18,Arrazola19,King09,Gromski19,Zwolak21}. We highlight that core to these tasks is the exchange of data through application programming interfaces (APIs). 
\subsubsection{Data acquisition}
\label{sec:data}

Quantum hardware for control and readout of quantum systems includes a user interface to operate the instrument, for instance, through a touch-based front panel or remotely accessible GUI. However, to implement more complex tasks such as characterization, tuning, calibration and general operation of QPU elements that involve coordination and timing between several different pieces of programmable hardware, it is essential to add a software layer that runs on a control computer and communicates with hardware directly through a programming interface. This software layer orchestrates the flow of control and data acquisition commands and is responsible for the overall operation of the QPU via the quantum hardware.

\begin{figure}
    \centering
    \includegraphics[width=0.5\textwidth]{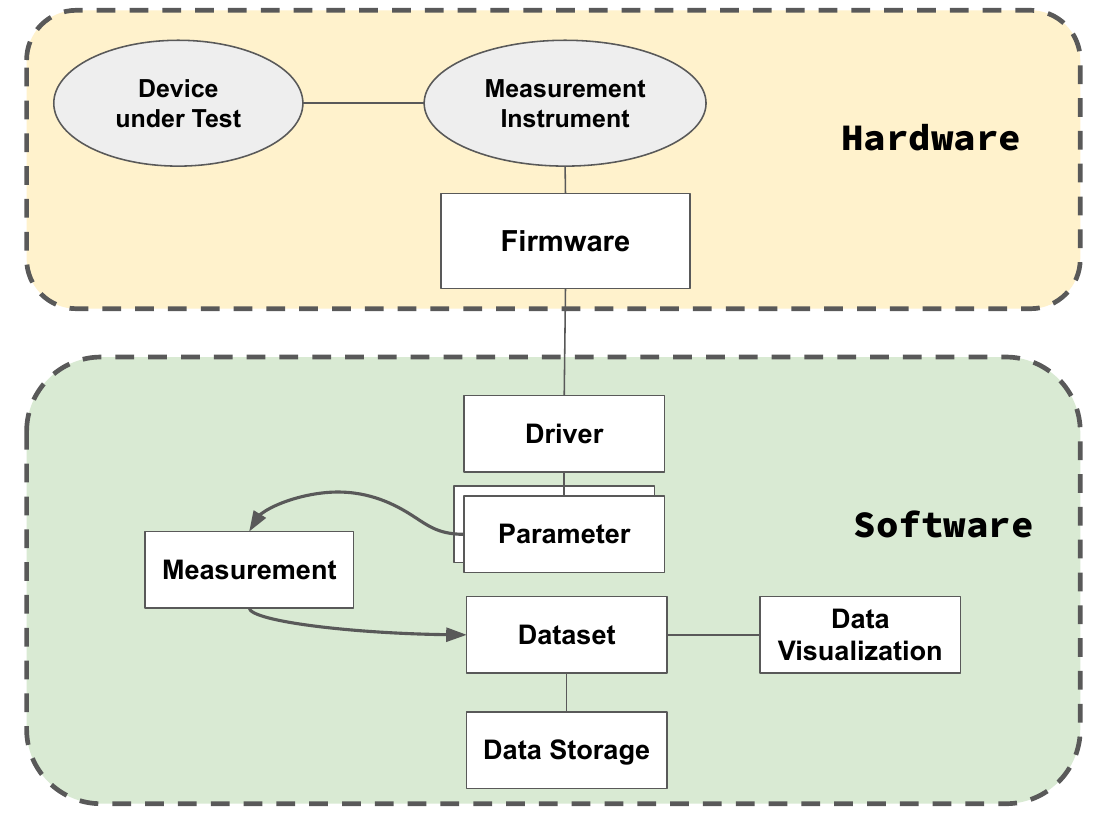}
    \caption{\textbf{Control and data acquisition.} The diagram shows some of the most important elements in the software and hardware stack involved in a quantum experiment. The hardware part of the stack includes the device under test or operation, the measurement instrument and the firmware. A driver provides interaction between the software and hardware. Drivers implement parameters that can be controlled (set) or read (get) over several measurements. The result of each measurement is stored into a dataset, that is read for data analysis, visualization and storage. }
    \label{fig:hwsw}
\end{figure}

On the software side, there exist quite a few alternatives for QPU control and data acquisition, starting with QCoDeS~\cite{jens_hedegaard_nielsen_2023_8344579}, which is a Python framework developed by Microsoft Quantum. Another example of Control and Data acquisition software is ARTIQ~\cite{Bourdeauducq_2016}, which is also used to control hardware, as discussed in Sec.~\ref{sec:control}. Control and data acquisition software encompasses various elements and instruments: the quantum device, the firmware that interconnects with the instrument drivers, parameters that are controlled and measured to generate datasets, which are stored, analyzed and explored via data visualization tools, as depicted in Fig. \ref{fig:hwsw}.

To achieve frictionless control of a QPU, the software typically abstracts away the hardware elements using drivers. For example in the QCoDeS library~\cite{jens_hedegaard_nielsen_2023_8344579}, an instrument driver is a Python class that implements methods to operate the hardware using commands that are sent to the device via the programming interface. Namely, a quantum hardware instrument, such as an arbitrary waveform generator, can be used to generate a radio frequency pulse for a rotation of a qubit, e.g., an X gate. In order to program this pulse, the driver implements parameters such as frequency, amplitude, phase and pulse time. The driver will then use these values to construct the right programming commands to send to the instrument to set up the pulse. The QCoDeS documentation~\cite{jens_hedegaard_nielsen_2023_8344579} contains driver examples to control over 50 instruments.

The data acquisition sofware uses instrument drivers to form an abstraction layer to the several pieces of hardware that control a QPU. Parameters can also be used in higher levels of abstraction on top of instrument parameters to represent QPU device elements that are controlled and read via the hardware layer, such as a qubit readout circuit. The parameters are then used in automated measurement routines with the goal to characterize, calibrate or program a QPU or its different elements. This requires saving, analyzing and visualizing the data that is recorded by these instruments in a dataset. Most data acquisition software frameworks therefore also include data storage abstractions and data visualization tools.
In order to control the QPU elements using quantum programming instructions, the control signals need to be calibrated for optimal gate and read-out fidelity. As shown in Fig. \ref{fig:overarching-diagram}, calibration values are obtained via device testing and calibration measurements using the data acquisition software, analysis routines and data storage in the Cloud. More details can be found in section \ref{sec:control} on optimal control, calibration and characterization for high-fidelity quantum operations.

A way to approach the software-hardware interface is through an Instruction Set Architecture (ISA)~\cite{isa} as it is done in classical computing. The ISA acts as a contract between hardware and software by explicitly describing the set of available operations, machine instruction format, supported data types, registers, and memory addressing. A well-designed ISA can present a very compact and efficient method to access the specialized features of a computing device. Graphical Processor Units (GPUs)~\cite{Lindholm:2008:NTU:1373105.1373197} and their success in becoming an integral part of many modern devices is an example of instruction set design and standardization leading to widespread specialized architecture adoption. 

There are a few quantum ISA implementations~\cite{balensiefer2005evaluation,smith2017,fu2019eqasm}, with some that explore this control and data acquisition approach, such as QUASAR~\cite{9325421} developed at Advanced Quantum Testbed (AQT), Lawrence Berkeley National Laboratory (LBNL). QUASAR is based on the RISC-V ISA~\cite{Waterman:EECS-2016-129} - an open-source architecture that revolutionized the classical computing field and continues expanding toward emerging technology applications. QUASAR has been demonstrated in conjunction with QubiC~\cite{Xu21} (QubiC is discussed separately in section \ref{sec:hardware}, below), executing experiments, including mid-circuit measurement and feed-forward, on superconducting quantum processors (QPU) at AQT, LBNL.

\begin{myframe}[width=\columnwidth]
\emph{Example 2, Control and Data Acquisition: }%\newline
{\bf ``QUASAR: A Quantum Instruction Set Architecture.'' }%\newline 
The QUASAR development started in 2017 and went through several iterations. It is a project defining a quantum Instruction Set Architecture (ISA) to interface software with hardware. The first version was a tightly-coupled extension that required significant changes to the processor micro-architecture. Later, the implementation moved towards a more decoupled approach.
The current implementation is a RISC-V Rocket Core~\cite{rocket} extended with the QUASAR co-processor (note that RISC-V is a very popular open standard ISA in classical open hardware. RISC-V is based on the established reduced instruction set computer (RISC) principles, and the Rocket Core is a processor that implements RISC-V.)
In contrast to classical instruction sets, quantum ISAs operate on different types of computations. Traditional general-purpose operations are supplemented with a set of basic quantum gates applied on direct-addressed qubits and/or qubit registers. The RISC-V core executes the main program; when a quantum instruction occurs at the fetch pipeline stage, it forwards it via the RoCC interface~\cite{chipyard} to be decoded and executed by the QUASAR co-processor. Such an architectural solution allows the main core to continue performing computations while the quantum backend is generating the control pulses. Moreover, the RISC-V core can perform complex computations, such as floating-point arithmetic for phase estimation, and make conditional branching during algorithm execution based on the qubit measurement data. That makes the QUASAR implementation flexible enough to accommodate complex hybrid experiments for classical-quantum computations. The RISC-V Rocket core runs a Linux kernel that facilitates remote communication with the user host machine and provides additional functionalities for the quantum software stack.

\end{myframe}

\subsubsection{Pulse-level control with hardware integration}
\label{sec:hardware}
In recent years, several pulse-level control software packages have been developed with the idea of providing end-users with a tool to program all the relevant device-specific physical parameters of the system~\cite{Bourdeauducq_2016,Stefanazzi21,Alexander_2020,Silverio2022,Li2022,lobser2023jaqalpaw}. This approach allows for a finer level of control over pulses during the application of gates, and also makes it possible to directly use the Hamiltonian of the system as a resource for computation. Within the gate-level framework of quantum circuits, pulse-level control and simulation enables a greater level of flexibility and the ability to implement optimal control schemes. Within the analog quantum simulation framework, pulse-level control notably allows practitioners to take advantage of the mimetic capabilities of the hardware. Exposing the hardware controls at such a low level is intended to help quantum developers design software procedures while having the specific characteristics of the hardware in mind~\cite{tsai2022}. 

Control hardware, e.g., field programmable gate arrays (FPGAs), arbitrary waveform generators (AWGs), sequencers, have historically been closed-source, either provided by industrial manufacturers or by in-house developed systems. 
The ARTIQ/Sinara project is a notable exception, focused on the fast control of ion-based quantum processors through FPGAs, as detailed in the box below (\emph{Example 3}). 
\begin{myframe}[width=\columnwidth]
\emph{Example 3, Control and Data Acquisition: }%\newline
{\bf ``ARTIQ: Advanced Real-Time Infrastructure for Quantum physics.'' }%\newline 
The ARTIQ experiment control and data acquisition system was initiated in 2013 at the NIST Ion Storage Group, in partnership with M-Labs Ltd, to address the deficiencies observed with control systems developed in-house by physicists or based on existing commercial solutions.
A key feature of the ARTIQ system is a high-level Python-based programming language that helps describe complex experiments, which is compiled and executed on dedicated FPGA hardware with nanosecond timing resolution and sub-microsecond latency. Using the abstractions provided by Python, ARTIQ has the capability to handle the entire control stack from quantum circuits to pulse-level programming. ARTIQ also supports connecting several FPGAs together and synchronizing their clocks in the sub-nanosecond regime, which greatly expands the input/output (I/O) scalability of the system.

	Initial versions of the ARTIQ system ran on physicist-designed hardware based on FPGA development kits with custom I/O expansion boards. In order to improve the quality, availability and reproducibility of the hardware, the Sinara project~\cite{Kulik_2022} was started in collaboration with Warsaw University of Technology. The Sinara project developed a modular system with carrier FPGA cards (Kasli, Kasli-SoC) controlling various so-called Eurocard Extension Modules (EEMs) catering to the needs of each experiment -- such as digital I/O, Analog to Digital Converter (ADC), Digital to Analog Converter (DAC), Direct Digital Synthesis (DDS), Phase-Locked Loop (PLL) synthesizer, AWG. 
 There is ongoing work by Duke University on firmware~\cite{Duke-artiq} and by Warsaw
University of Technology on hardware~\cite{Sinara} to port ARTIQ to the RF-SoC platform, with similar capabilities to QICK.

 Combined with the ARTIQ software, the Sinara hardware has encountered substantial success, with almost a thousand quantum physics experiments relying on ARTIQ/Sinara systems.
	
  %The ARTIQ and Sinara authors, contributors, and supporters consider the free and open exchange of scientific tools to be equally important and have chosen the licensing terms of ARTIQ and Sinara accordingly. 
ARTIQ, including its gateware, the firmware, and the ARTIQ tools and libraries are licensed as LGPLv3+. The Sinara hardware designs are licensed under CERN OHL. This ensures that a user of ARTIQ or Sinara hardware designs obtains broad rights to use, redistribute, study, and modify them.
\end{myframe}

More recently, other FPGA-based projects have emerged beyond ARTIQ, for SC qubits control, such as QubiC~\cite{Xu2022} and QICK~\cite{Stefanazzi21}, two projects based on software, firmware and hardware composed of radio-frequency system-on-chip (RF-SoC) boards. 
Open-source FPGA-based control efforts for SC qubits started in 2018 with QubiC, developed at the Advanced Quantum Testbed (AQT) based at LBNL (more information on this facility is given in Sec.~\ref{sec:testb}, discussing Testbeds). The first version of QubiC was implemented on the XilinX VC707 platform ~\cite{huang2019scalable,Xu21} informed by actual quantum information science experiments at AQT. To reduce cost, complexity, and space requirements, customized hardware components were designed and fabricated, such as in-phase and quadrature (I/Q) mixing modules integrating filters, amplifiers, and bias\_Ts on printed circuit boards (PCBs) with Electromagnetic Interference (EMI) shielding, the design of which was published and open-sourced to benefit the community \cite{xu2021radio}. Additionally, automated single- and two-qubit gate calibration protocols were developed using QubiC~\cite{Xu2022}.

\begin{figure}
    \centering
    \includegraphics[width=1\textwidth]{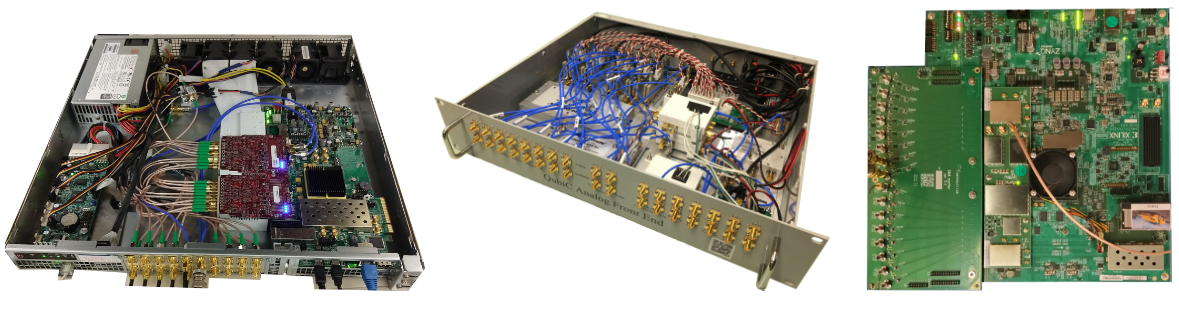}
    \caption{\textbf{QubiC, through the years:} QubiC 1.0, implemented initially on XilinX VC707, in its chassis with auxiliary components (left), the Analog Front End chassis for QubiC 1.0 (middle), and QubiC 2.0 \cite{xu2023qubic} implemented on XilinX ZCU216 with a custom SMA fanout board. A customized Surface Mount Assembly (SMA) fan-out board has been designed and recently fabricated to fully utilize all the channels of the ZCU216 board. Separately, a low-cost DAC extension was developed for the VC707 platform, to meet the varying frequency needs of different superconducting QPU architectures. The QubiC team has recently demonstrated heterogeneous synchronization of two sets of VC707 boards with low-cost DACs and two ZCU216 platforms \cite{xu2023fpga}.
}
    \label{fig:QubiC}
\end{figure}

To take advantage of the growing capabilities of recent RF-SoC platforms, QubiC was then ported to the XilinX ZCU216 platform \cite{xu2023fpga}, capable of direct generation of control pulses at higher frequencies (up to 10GHz, in the second Nyquist zone). Additionally, a novel Distributed Processor design was created and implemented to enable distributed decision-making and branching, such as mid-circuit measurement and feed-forward \cite{fruitwala2023distributed}.
The different iterations of QubiC 1.0 and QubiC 2.0 \cite{xu2023qubic} implementations are shown in Figure \ref{fig:QubiC}.
To form a prototype full-stack quantum control system with increased potential for future experiments, QubiC and QUASAR (described above in \ref{sec:data}) have been integrated and applied to quantum computing experiments at AQT, LBNL \cite{butko2023real}.

\begin{figure}
    \centering
    \includegraphics[width=0.5\textwidth]{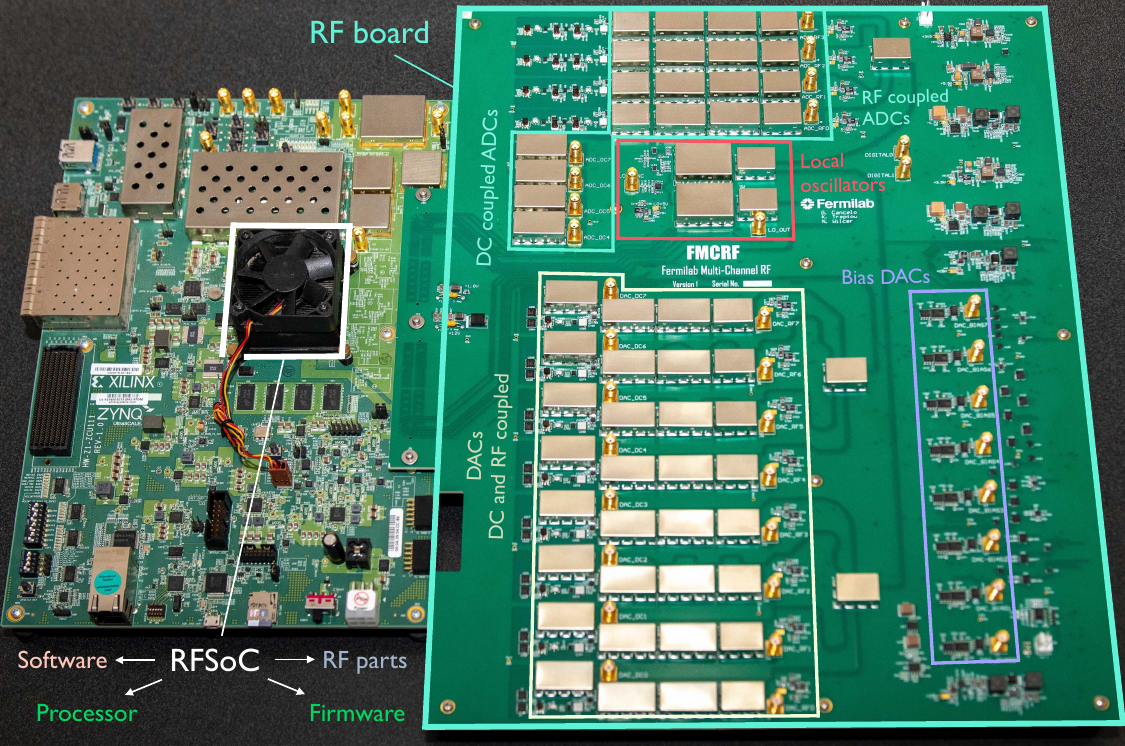}
    \caption{\textbf{The Quantum Instrumentation Control Kit (QICK).} The QICK consists of two pieces of hardware: a commercial RFSoC evaluation board (left), which connects to the QICK RF\&DC custom board (right) which can be used for additional signal amplification and filtering. QICK supports the AMD-Xilinx ZCU111~\cite{ZCU111} (shown in the picture with its companion custom board) and the AMD-Xilinx ZCU216~\cite{ZCU216} (not shown) with a new companion board under fabrication and testing.
}
    \label{fig:qick}
\end{figure}

Boards from the QICK project have reached over 40 labs in the USA and abroad after a little more than two years. Using QICK helps reduce equipment costs and increases the performance of quantum information science experiments with SC qubits~\cite{johnson2023exploration}. All of QICK implementations have been on Zynq boards, with the firmware provided by PYNQ (Python for Zynq), compatible with multiple generations of RF-SoC FPGAs. An image of a QICK board is shown in Figure~\ref{fig:qick}.
The usage of QICK has already expanded from SC qubits to atomic, molecular, and optical physics qubits and spin qubits (Nitrogen-Vacancy-center qubits), on the hardware side. On the application side, QICK is used not only for quantum computing but also for 
quantum sensing experiments, e.g., for dark matter candidates detection~\cite{Dixit_2021}, as well as for RF-SoC control in particle physics detection beyond quantum technology.

\subsubsection{Pulse-level simulation}
\label{sec:pulse}
 Simulation is important to design an experiment before running it on hardware and then to validate and interpret the results after data collection. Pulse-level simulation can be employed for quantum optics experiments, quantum simulation, and both for gate-level (digital) quantum computing and  analog quantum computing. In quantum computing, although gate-level instructions are the most common formalism on the user-side to write quantum algorithms, there can be a compiler that transforms discrete operations into pulses. 
Examples of pulse-level simulators include Pulser~\cite{Silverio2022} and Bloqade.jl, two toolboxes for neutral-atom QPUs respectively developed in Python and Julia. qutip-qip is a simulation toolkit that provides the freedom to design any QPU and simulate pulse level execution, integrating it QuTiP's noisy time evolution~\cite{Li2022}. We summarize information on Pulser in \emph{Example 4}. 
\newline
\begin{myframe}[width=\columnwidth]
\emph{Example 4, Pulse Level Simulation: }%\newline
    {\bf ``Pulser: A library for pulse-level/analog control of neutral atom devices.'' }%\newline
The Pulser framework is an open-source software library for designing pulse sequences for neutral-atom QPUs~\cite{Silverio_2022}. In such devices, individual atoms are placed in arrays of microscopic traps~\cite{Browaeys2020,Henriet2020,Nguyen23} and their quantum state is manipulated through the application of laser or microwave pulses. Using Pulser, developers can control all the relevant physical parameters of the qubit register and the driving channels. 

The central object in Pulser is called a sequence, which is linked to a device. The device contains information about the hardware constraints, such as the maximal number of qubits available, the minimal pairwise distance between qubits which can be reached or the number of lasers and the maximal amount of laser power available for each of them. These constraints are enforced upon the register, where the neutral-atom array is defined, and upon the pulses that are added to the sequence. Each pulse is defined by its phase, amplitude and detuning waveforms, and pulses are sequentially added to the channels that are available on the device. The resulting program can subsequently be sent to QPUs and executed on the hardware after a device-specific calibration-aware compilation step. In addition to providing an interface to hardware, Pulser includes a built-in emulator relying on QuTiP~\cite{Johansson_2012} that faithfully reproduces the hardware behavior. 
To this extent Pulser acts also as a simulator for QPU design. 
\end{myframe}
\subsubsection{Optimal control, calibration and characterization}
\label{sec:control}
High fidelity operations require high fidelity optimized controls. The process of generating optimal pulses traditionally involves a model-based open-loop optimal control step (in simulation) such as GRAPE, Krotov, GOAT algorithms~\cite{werschnik2007quantum,koch2022quantum,Goerz_2019} followed by a model-free closed-loop calibration step (in hardware). The former relies on building a sophisticated model of the system and the optimality of the generated pulses is constrained by the model’s capability to faithfully reproduce the imperfections of open quantum systems. The latter usually involves the use of a gradient-free optimization algorithm, e.g., Nelder-Mead or CMA-ES to optimize the parameters of an open-loop optimal pulse on the quantum hardware benchmarked with a fidelity measure such as ORBIT~\cite{Kelly2014ORBIT}.  
Open-source libraries such as JuliaQuantumControl, QuOCS and C3-Toolset~\cite{Wittler_2021,roy2022software} aim to provide a broad range of the optimal control functionalities previously discussed, with easy to use interfaces in Python or Julia. These tools also incorporate automatic differentiation (AD) techniques which have made it possible to obtain gradients of arbitrary order for free, even with the most complex numerical simulations. Such gradients are essential for the optimization of pulse parameters in open-loop control. AD capabilities are provided by the use of standard Machine Learning frameworks such as TensorFlow or JAX~\cite{Tensorflow16,Frostig2018}.
\newline
\begin{myframe}[width=\columnwidth]
\emph{Example 5, Control \& Data Acquisition: }%\newline
    {\bf ``Qudi: A modular python suite for experiment control and data processing.'' }%\newline
Qudi provides a flexible and modular framework for controlling and analysing data from confocal microscopy, quantum optics and quantum information experiments primarily based on color centers in diamond. Developed using Python and Qt, it is a feature rich open-source solution providing a GUI, real time data visualization, distributed execution over networks, Jupyter notebook interface, configuration management, and data recording \cite{Qudi_Softwarex_2017}. A key feature of Qudi is its modular design consisting of functionally independent parts that are orchestrated by a manager component with a declarative configuration management system. This also makes it very easy to extend the framework to support new devices and experiments. The entry point for any user is the start.py file which loads the core components that take care of configuration parsing, module management, error logging, and remote network access. There is a three-tier architecture consisting of logic modules, hardware modules and GUI modules ensuring a clear separation of concern and allowing independent use, development and deployment.

\end{myframe}

In order to improve the fidelity of the pulse obtained through open-loop control, both the model parameters and the model itself need refinement. One approach for refining model parameters is the technique of data-driven system identification also known as characterisation. This is achieved through learning model parameters from data collected during experiments performed on the hardware, e.g., the calibration step previously outlined. Refining the model translates to building a complex digital twin that models not only the quantum dynamics but also all of accessory classical electronics and their non-ideal behavior. These two techniques — Model Learning and Quantum-Classical Digital Twin — are tightly integrated as a unified solution in the C3-Toolset package~\cite{Wittler_2021}, as discussed below (\emph{Example 5}). 
\newline
\begin{myframe}[width=\columnwidth]
\emph{Example 6. Optimal Control, Calibration and Characterization: }\newline
{\bf ``C3: An integrated tool-set for Control, Calibration and Characterization.''}\newline
The C3-Toolset package provides an API for directly interfacing with hardware to use the pulses generated by optimal control and further optimize them using closed-loop calibration algorithms. Users have access to a high-level Qiskit interface that allows them to run gate or pulse level quantum circuits on the full-physics noisy high-fidelity differentiable simulator. Libraries like QuOCS \cite{quocs} also provide an extensive suite of model-based optimal control and pulse-shaping for a wide variety of hardware platforms.
In recent years, a variety of machine learning for quantum control techniques have been closely integrated in the quantum device bring up process. Besides the previously mentioned process of learning model parameters, reinforcement learning (RL) has been particularly useful in a variety of applications. The open-source library rbqoc~\cite{propson2022robust} details a technique for noise-robust control using RL.

Calibration is generally needed for most platforms. Depending on the architecture, this involves different control signals. A difference between academic research labs and industry is the tendency to automate and standardize calibrations tasks. Tests are run periodically to validate the status of the system before running an experiment and during an experiment. On a schematic level, these can be seen as continuous integration (CI) tests, involving hardware control but similar to open-source testing for software projects. If the job does not pass a test, automatic re-calibration is prompted. 
\end{myframe}

%\subsubsection{Pulse-level simulation}
%Simulation is important to design an experiment before running it on hardware and then to validate and interpret the results. Pulse-level simulation is important both for gate-level quantum computing and for analog quantum simulation and quantum computing schemes. Although gate-level instructions are the most common formalism on the user-side to write quantum algorithms, there is always a compiler that transforms discrete operations into pulses. Within the gate-level framework of quantum circuits, pulse-level simulation enables a greater level of control and the ability to implement optimal control schemes. Within the analog quantum simulation framework, pulse-level simulation allows to take advantage of the mimetic capabilities of the hardware. 

\subsection{Designs for quantum error correction}
\label{sec:qec}
Many applications of quantum computing, networking, and sensing technologies will require a significant reduction in error rates. In the long term, developing quantum hardware with built in error correction can address this need through fault-tolerant quantum computing. In this section, we review some of the open-source software available today for studying and simulating quantum error correcting codes. This software is becoming increasingly important as quantum error correction moves from theory into practice.

Quantum error correcting codes are often based on stabilizer subsets of quantum computation and so can benefit from specialized simulators. A leading open-source package for stabilizer simulation today is Stim~\cite{gidney2021stim} which improves on existing stabilizer simulators available in Qiskit, Cirq, or other packages.  

Other libraries sit on top of simulators like stim and allow users to explore quantum error correcting code designs. These packages help inform longer term planning for quantum processor architecture. Examples include plaquette~\cite{qc-design2023} and stac~\cite{khalid2023}. Both of these are Python libraries that have built in examples of quantum error correcting codes and allow you to generate logical circuits to study performance. There are also tools that act as compilers, translating logical circuits into physical circuits that can be run directly on capable hardware, e.g. the suite of tools developed at latticesurgery.com for compiling to the surface code. 

Finally, there are open implementations of the decoding algorithms used to decide how to best correct error detected by different quantum codes. PyMatching~\cite{higgott2021pymatching, higgott2023sparse} is an example open-source decoder.
This software can help inform the design of processors specifically adapted to run quantum error correction. For example, stim was used to simulate novel superconducting qubit designs to support surface code parity measurements in~\cite{reagor2022hardware}.

There remain many important opportunities to improve quantum error correction. For example, fast classical control of QPUs is essential for feedback loops in quantum error correction implementation. Tailored tools could reduce the existing gap between hardware operation and software instructions. More generally, increasing the ecosystem of open-source software available in this field will both support new breakthrough ideas as well as accelerate the transition from theory to practice in building fault-tolerant quantum computers.

\subsection{Open quantum hardware facilities}
\label{sec:facilities}
In these section we review the state of the art with respect to facilities that can enable a robust OQH ecosystem. We find three qualitatively different categories: open-access to QPUs and remotely accessible research labs, collaborative testbeds, and facilities for fabrication such as foundries. 

%\subsubsection{Cloud access to QPUs by industry  providers}
%Currently, the most widespread open-access to QPUs is provided by cloud infrastructure provided by industry  players such as corporates or startups. Some eminent examples include the Rigetti Cloud and the IBM Quantum Experience, which both pioneered open-access to their own devices. In a second stage of the industry development, there has been a partial differentiation (with possibly overlapping roles) between cloud providers, such as AWS Braket and Azure Quantum and QPU providers, such as startups like Quantinuum, IonQ, and so on. Some startups provide open access through their cloud, such as Quandela or IonQ. Over the years, the level of control over device properties in compilation, optimization, qubit mapping, etc., has increased, spilling notably into remote pulse level control, both for digital (Qiskit pulse~\cite{Alexander_2020} )and analog quantum computing (Pasqal's Pulse studio~\cite{Henriet2020}). This list is by no means comprehensive as several QPU providers have emerged in the space. We believe that open access from the first industry providers can usher an era of even more widespread access to devices, as addressed in the following sections.  
\subsubsection{Remotely accessible labs and cloud-connected labs}
\label{sec:remotelabs}

The idea of open remote labs, i.e., laboratories that can be remotely accessed by users to perform real experiments is not new in the context of scientific research and education~\cite{saenz2015open, gomes2009current}. 
For example, the educational tool HYPATIA~\cite{kourkoumelis2014hypatia}  can be used to perform particle-physics experiments with real data produced by the ATLAS experiment at CERN. In a similar way, quantum computers can be used as quantum remote labs,  i.e., advanced laboratories in which quantum experiments can be performed for purely scientific purposes without any computational motivation~\cite{Norman18,DuBois20,Bussmann21,Clark21,Altman21,Awschalom21,Alexeev21,Frey21}. For example in Ref.~\cite{solfanelli2021experimental} a quantum computer was used for testing quantum fluctuation theorems, while in Ref.~\cite{mi2022time} a quantum computer was used to prepare a many-body quantum system in a time-crystalline phase.

Superconducting-circuit quantum computers originally became available online from IBM Quantum with the IBM Quantum Experience, Rigetti Computing with the Rigetti Quantum Cloud Services, and other providers. These have had an impact on the way research is done in quantum computing and quantum optics as well as enabling access for students for outreach activities, due to the fact that experiments can be done much more easily (even by theorists), from the cloud. We are now in a second phase with more providers putting their devices online, encompassing more technologies – ion-based (IonQ, Quantinuum), atom-based (e.g., Infleqtion, Pasqal, QuEra), photonics-based (e.g., Quandela, Xanadu), quantum-dot based, etc. 

Turning the attention to institutional frameworks, the European OpenSuperQ project (and the follow on OpenSuperQPlus project) \cite{opensuperq} is similarly aimed at developing public-owned full-stack open-access superconducting systems. Multiple remotely accessible QPUs are online or currently being set up as part of this project at the Delft University of Technology, the Wallenberg Centre for Quantum Technology (Chalmers), the Walther-Meissner-Institut (Munich), and at the Forschungszentrum Juelich (from phase 1). A hybrid form of cloud access is also emerging, in which cloud providers give access to different quantum processor providers. Further sustaining open access to QPU providers can be important for scientific discovery.

 Over the years, the level of control over device properties in compilation, optimization, qubit mapping, etc., has increased, spilling notably into remote pulse level control, both for digital (e.g., Qiskit pulse~\cite{Alexander_2020}) and analog quantum computing (e.g., Pasqal's Pulse studio~\cite{Henriet2020}). The deeper the access, the more advanced the control researchers have over hardware and the greater the opportunity for integration of open-hardware features with cloud-access.
An example of novel interaction between cloud providers and OQH is given by the Amazon AWS application developed for QICK, called SideQICK, which is being integrated into a more general cloud queue for quantum devices~\cite{cloudqueue}.

We also witness examples of a hybrid model of industry-public interaction enabling open access. Quantum computers are being shipped to research centers and further integrated with HPC centers, which also act as simulator infrastructure (such as the EU HPCQS). This provides new avenues for industry-academic collaborations and purpose-specific hardware customization, use cases and access. 
It will be important to foster academic research labs putting their quantum devices online, building upon the open-source tools described in the previous sections. This could further change the research landscape, enabling more researchers to test research ideas on more devices, novel architectures and platforms, and in turn facilitate technology transfer.

\subsubsection{Deeply collaboratively testbeds}
\label{sec:testb}
While cloud-based platforms allow scientists and the public alike to submit their circuits and receive the results, deeper, customized access to the full stack to enable more involved experiments and R\&D efforts is generally not possible on these platforms. To provide such access to users in Academia, Industry, and National Labs, the US Department of Energy has funded two testbed programs: QSCOUT (Quantum Scientific Computing Open User Testbed), based on trapped ions and located at Sandia National Laboratories, and AQT (Advanced Quantum Testbed), based on superconducting qubits and located at Lawrence Berkeley National Laboratory. A similar network called Qu-Test \cite{qupilot_qutest} managed by TNO at Delft, Netherlands has recently been kicked off with the goal of providing a federated network of testbeds by bringing together 13 service providers and 11 industrial users from the European quantum community. Likewise, in the UK, the National Quantum Computing Centre (NQCC) is building, testing and hosting full stack quantum computing solutions \cite{nqcc}. Through close collaboration between industry, government and academia, the NQCC is initially focusing on developing demonstrator platforms which provide direct access to quantum computing resources while helping to also develop a robust supply chain. A recently concluded funding call by the Innovate UK \cite{sbri_testbed} earmarked 30 million GBP to build, commission, and validate an operational quantum computing testbed in the UK as part of the NQCC.

In Canada, the Open Quantum Design (OQD)initiative aims at building a completely open-source and open-hardware quantum computer based on ions. OQD has taken the form of a full-fledged non-profit organization arising from previous experimental and theoretical work performed at the Institute for Quantum Computing and Perimeter Institute.

These platforms allow low-level access to the full stack of quantum hardware (including programming languages~\cite{Morrison,jaqal}, gate-pulse shaping~\cite{lobser2023jaqalpaw}, noise injection, unique qubit/qudit states, specialized calibrations, comparison of compilation techniques, etc.) allowing users to probe how the machines actually work and how to make them work better. The testbeds foster deep collaborations between the host laboratory and the users.

Collaborations on these testbeds have so far resulted in demonstrations of scientific applications of near-term quantum computers, and the development of many tools to benchmark quantum computers as well as to characterize and mitigate errors on them QPUs. This could further change the research landscape, enabling more researchers to test research ideas on more devices.  Moreover, the existence of testbeds would be crucial to enable reproducible benchmarks, for quantum hardware and for application-driven tasks.  Testbeds are key to facilitate startup creation in the quantum space, as they lower the barrier for testing prototypes with expensive equipment (such as dilution fridges).

\subsubsection{Fabrication and foundries}
\label{sec:foundr}
As described in Section \ref{sec:HWdesign}, open-source tools exist to facilitate the design and creation of quantum processors \emph{in silico}. However, translating those designs into actual hardware can be difficult and expensive. Current arrangements for producing hardware on the basis of designs fall into four categories: partnerships, the construction of research-grade fab facilities, the use of academic foundries, and ``quantum-fab-as-a-service". 

Several companies in recent years have begun partnerships with major manufacturers to procure supply lines for their hardware designs. For example, photonic startups PsiQuantum and Xanadu have both signed agreements with semiconductor manufacturers GlobalFoundries~\cite{psiqgf,xanadugf}. Trapped-ion startup IonQ has sourced some of its traps from Sandia National Laboratories~\cite{Clark21}. These sorts of partnerships are typically only available to startups or other corporations, and allow for the use of use of existing manufacturing techniques and scalable processes. However, whether rapid prototype of hardware can be achieved is unclear.

An alternative approach is to stand up one's own fabrication facility. Startup Rigetti Computing is notable in this space for their ``Fab-1" facility~\cite{rigettifab1}, which allows the company to prototyping new hardware. However, such a facility can be expensive to construct, and requires access to large amounts of capital.
In the United States, in recognition of the need for fabrication facilities, various foundries are being stood up around the country, including the UCSB quantum foundry~\cite{ucsbfoundry}, MonArk quantum foundry~\cite{monarkfoundry}, and the LPS Qubit Collaboratory~\cite{lpscollaboratory}. The European Commission has established a somewhat similar program called Qu-Pilot \cite{qupilot_qutest} under the umbrella of the Quantum Flagship consisting of 21 partners from 9 different countries, with the goal of developing and providing access to federated European production facilities linking existing infrastructure. The overall coordination of the project is managed by VTT, Finland.

Finally, some startups have leveraged this need toward the development of new businesses. For example, startup Quantware in the Netherlands~\cite{quantware} helps designing, developing, and fabricating hardware for customers. In Canada, the NSERC CREATE programs have partnered with CMC Microsystems for novel superconducting circuits workshops at the Stewart Blusson Quantum Matter Institute. In the future, more industry players, industry consortia, startup incubators, and academic partnerships could enable even more facilities for processor fabrication, inspired by existing electronics industry frameworks, such as the Efabless~\cite{efabless} and the Google Silicon project~\cite{google-silicon}. 

\section{Discussion: Current Gaps \& Future Recommendations}
\label{sec:gaps}

In Sec. II we reviewed OQH today, with an overview of the various OQH categories and deep dives into selected projects. From this overview it is possible to draw an overarching picture and outline some major topics across the OQH ecosystem. In the sections below we list these topics, identifying gaps and making recommendations to close them.
\\

\noindent{\bf OQH growth and maintenance across technologies and architectures.}
%\subsubsection{OQH growth and maintenance across technologies and architectures}
Currently, most of the OQH projects focus on tools for quantum computers. There are considerable opportunities for expanding OQH projects and tools to accelerate scientific discovery and tech transfer in quantum communication, quantum metrology, and quantum sensing.
Moreover, given the early stage of the field, for specific functionalities, tooling supporting only a given set of architectures may be currently available. 
For example, the open hardware tools for chip design are mostly revolving around SC qubits (and partially ion traps). There is room for tools for other QPU architectures, such as atom-based, spin-based, and photonics-based QPUs.
With respect to control, open hardware projects including firmware and hardware, such as FPGAs, have been developed mostly for ion traps and SC qubits and can more broadly applied to other architectures. Moreover, while it is possible to find multiple open-source FPGA (and general open-hardware electronics projects with many being applicable to physics), one would be hard-pressed to find anything that comes closest to an optical frequency comb in photonics tooling. 
\\

\noindent{\bf APIs and standards for instruction sets.}
%\subsubsection{APIs and standards for instruction sets}
We note that interoperability is a major challenge in the OQH ecosystem. From APIs in the software stack to software-hardware integration via ISA, there is work to be done. Currently, there is considerable duplication on higher software stack and one needs to do conversions of quantum programs in order to run them on different QPUs. Frameworks such as the Quantum Intermediate Representation (QIR) project and OpenQASM (Open Quantum Assembly Language)~\cite{Cross_2022,Cross2017open} can help at the higher level. Broader support for common frameworks getting further down at the hardware control level would be desired, such as QUASAR and QUA.
\\

\noindent{\bf Benchmarks.}
%\subsubsection{Benchmarks}
Open-source benchmarking suites can have strong impact on the scientific community, providing information on the state of the art of current hardware architectures. 
%Benchmarks should help convey how different platforms perform a task compared to themselves  -- in the same family of quantum devices --, and compared to others -- quantum vs. classical computing or quantum vs. other non-conventional computers, like Ising machines. 
So far we note that quantum technologies have been confronted with a lack of standardized benchmarks. Benchmarks are useful for evaluating the performance of quantum devices and algorithms, as well as assessing the relative performance compared to purely-classical solutions in relevant applications or non-conventional computers~\cite{ising}. Thankfully, efforts in the standardization of benchmarks, such as those provided by the QED-C, a U.S.-based industry consortium ~\cite{Lubinski2023,Lubinski2023optimization,Amico2023} have been occurring. 
Pipelines for crowd-sourced benchmark results is also needed: \url{https://metriq.info/} is one attempt, inspired by projects such as Papers With Code and MLCommons in machine learning. Pipelines including OSS and OQH can simplify the accomplishment of standardized benchmarks.  
\\

\noindent{\bf Open access to hardware.}
Currently, there exist several cloud-accessible QPUs and cloud service frameworks. Providers have made QPUs accessible in a variety of ways, including direct and free access, as well as through partnerships within institutional frameworks (e.g.,ORNL's Quantum Computing User Program (QCUP) for national labs and federally-funded consortia). However, the research community would benefit from further access enabled by grants and other awards. Ensuring  such frameworks and access points increase as soon as possible is necessary for creating a more equitable research landscape.
Moreover, while several research labs have infrastructure to enable remote access and data acquisition of experiments, this is generally for internal use. Sometimes, access is extended to partnering organizations or collaborating researchers. However, there is a general lack of research labs who have put their device online for cloud access to a wider community of users and researchers. This is due to the overhead in infrastructure involved in making such devices operational and maintained. OQH can help with this regard, by providing researchers with existing tools that can be plugged in to bring up experiments from research labs.
\\

\noindent{\bf Reproducibility.} In order to facilitate reproducibility of results, journals and academic guidelines can further encourage (and mandate where appropriate) the sharing of open software along with research publications. Funding agencies, universities and other research stakeholders can facilitate these efforts by recognizing software artifacts as important, citable outcomes for measuring research impact. 
\\

\noindent{\bf OQH policy and intellectual property.}
%\subsubsection{OQH policy and intellectual property}
Implementing policy that is open-hardware aware can favor the growth of a OQH ecosystem. For example, learning from existing open hardware projects beyond quantum technology can help avoid known pitfalls. A practical example is the adoption of an integrated toolchain fostering collaboration between academic labs, software developers, and industry. An example at the facility level is the investment into the establishment of shared facilities, such as quantum foundries, which can provide a flywheel effect for startups. 
Policy makers and quantum experts have the opportunity to work together, in order to ensure that informed decisions are made about the application of export control and intellectual property protection laws to quantum hardware. On the one hand, this could ensure that policy makers provide clear guidance about how efforts to open up quantum hardware will be treated from the perspective of these laws. On the other hand, researchers can develop expertise on the correct use of licenses for open-source software and open hardware. This guidance can include implementing processes through which one can more easily assess what hardware-based technology should be open-sourced, and what should not, making strategic calls at a narrow level (researcher, company) and wider level (common good for society). Researchers need to balance against intellectual property (IP) \& strategic competitive considerations for certain components and technologies that could provide commercial exploitation. At the same time, they can consider when OSS can be used in business models, helping grow a community of users and creators around products.
\\

\noindent{\bf Support of OQH in education and academia.}
%\subsubsection{Support of OQH in education and academia}
We believe it is important to foster OSS and OQH adoption in education and research.
One actionable item is to encourage schools and universities to include in course material and labs the usage of open-source software~\cite{dawes2019undergraduate} and open hardware~\cite{Bouquet_2017,Gingl_2019,Ulmann_2021}. Programs that prepare the future quantum workforce are fundamental to broaden the acquisition of talents and develop new career paths~\cite{Aiello21,Asfaw22,Dzurak22}.
This includes developing communities across domain knowledge.

A bottom-up approach to strengthening OQH is by encouraging academic labs to pull together their systems with existing open-source tools, so they can create do-it-yourself small-scale testbeds. If open-source tools are not available, researchers can consider developing and open-sourcing new tools or using less-closed tools (e.g., proprietary but with a layer of open APIs). Fostering the creation of new open-hardware projects includes engaging in activities, programs and material that facilitates the open-sourcing of existing projects, such as Ref.~\cite{shammah-code}, which provides information about starting a scientific OSS project. Starting an OQH project includes general challenges found by other OSS projects (testing, documentation, distribution, maintenance, community relations) and some specific ones, such as how to distribute physical devices to users, or work with experimental labs on ensuring device control. 

An example of a top-down approach to supporting open-source projects (and open hardware) is provided by funding agencies programs acknowledging the impact of OSS projects in science. One notable example is the NSF POSE program. Further tailored funding in programs focused in OQH could strengthen the ecosystem addressing some of its characteristics.
\\

\noindent{\bf Community building: projects support and governance.}
%\subsubsection{Community building: projects support and governance}
 Similarly to other OSS projects, also OQH projects need to interact with their community of users, engage feedback, create guidelines to foster code contributions and meetings. They can benefit from non-profit organization support for their governance, as typically scientific OSS projects start in a research lab or group and then grow in scope and need for representation of more parties and partners. 
 There is a need to simplify the bureaucracy in compliance at institutional at government level to contribute to open-source projects and open hardware, e.g., by equipping national labs with default policies that would allow researchers to contribute to open-source projects with permissive licenses approved by the Open Source Initiative (OSI). 
 Often the originating organization such a national lab has restrictions that prevent an open-source project from growing independently. This highlights the need for alternatively supporting organizations to incubate and house open-source projects. Examples of non-profit organizations that do this in the classical space are NumFOCUS, the Linux Foundation, and the Open Source Hardware Association (OSHA). Unitary Fund is a 501(c)(3) non-profit organization that offers this support specifically to quantum technology projects.

%\begin{table}[t]
%\centering
%\def\arraystretch{2}% 
%\setlength\tabcolsep{.2cm}
%\begin{tabular}{p{0.2\columnwidth}p{0.4\columnwidth}p{0.3\columnwidth}}
%\hline%\\
%{\bf Topic}& {\bf Recommendation}& {\bf Action}  \\
%\hline
%
%OQH applications and architectures &Expand OQH ``horizontally" on several fields and architectures\\
%
%Interoperability&
%Strengthen OQH ``vertically" & \begin{itemize}
%\item Build APIs \item Agree on standards for instruction sets \end{itemize}\\
%Benchmarks & 
%Consolidate benchmarks of QPUs \\
%Access&
%Enable open access and reproducibility\\
%\hline

%\end{tabular}
%\caption{{\bf Open quantum hardware recommendations}.}
%\label{table:recs}
%\end{table}

\section{Conclusions}
\label{sec:concl}
Quantum computers -- and quantum devices in general -- provide a fascinating prospect to bridge a platform for fundamental science, technology transfer in engineering, and novelties in cloud computing and actuators. An open hardware approach in quantum technology  will preserve, and possibly expand, the pre-competitive space necessary to make ideas flourish and enable accessible benchmarks and evaluations that have already characterized quantum software.   

As we have illustrated, quantum technology provides unique challenges and features with regards to the usability and feasibility of open-hardware stack: from process design toolkit to foundries, from simulation software to control and data acquisition, from cloud infrastructure to testbeds. All these layers can benefit from collaborative innovation. 
Learning from lessons learned in the early days of quantum software, and from other open hardware verticals in conventional computing and tooling, as well as from a snapshot of current gaps in the quantum ecosystem, we hope that the quantum industry can efficiently implement innovation, build lively communities, and expand its workforce. 

\section{Acknowledgments}
\label{sec:acknw}
This work stemmed from a workshop~\cite{Shammah21ws} hosted at IEEE QCE 2021. The authors thank Gary Steele and Zlatko Minev for fruitful discussions. NS thanks the organizers of IEEE QCE 2021 for the opportunity to start the discussion of OQH which culminated in this manuscript, and many of the co-authors for their contributions at that workshop. This material was funded in part by the U.S. Department of Energy, Office of Science, Office of Advanced Scientific Computing Research Quantum Testbed Program. Sandia National Laboratories is a multimission laboratory managed and operated by National Technology \& Engineering Solutions of Sandia, LLC, a wholly owned subsidiary of Honeywell International Inc., for the U.S. Department of Energy’s National Nuclear Security Administration under contract DE-NA0003525.  This paper describes objective technical results and analysis. Any subjective views or opinions that might be expressed in the paper do not necessarily represent the views of the U.S. Department of Energy or the United States Government. SAND2023-09347O.

This work was supported by the U.S. Department of Energy, Office of Science, Office of Advanced Scientific Computing Research, Accelerated Research in Quantum Computing under Award Number DE-SC0020266. AM acknowledges support from the PNRR MUR project PE0000023-NQSTI. QubiC and QUASAR research was funded by the U.S. Department of Energy, Office of Science, Office of Advanced Scientific Computing Research Quantum Testbed Program under contract DE-AC02-05CH11231

\bibliography{references.bib}

\end{document}